\documentclass[notitlepage,twocolumn,prl,tightenlines,footinbib,superscriptaddress,showpacs,floatfix]{revtex4-2} 

\usepackage{amsmath,amssymb,amsfonts,latexsym}
\usepackage[colorlinks=true, linkcolor=blue, citecolor=blue]{hyperref}
\usepackage{bbm}
\usepackage[mathcal]{euscript}
\usepackage{graphicx}
\usepackage{epsfig}
\usepackage{color}
\usepackage[en-US]{datetime2}
\usepackage{psfrag}
\usepackage{txfonts,pxfonts,wasysym,}
\usepackage{pifont}
\usepackage{tikz}

\newcommand\sbullet[1][.7]{\mathbin{\vcenter{\hbox{\scalebox{#1}{$\bullet$}}}}}
\newcommand{\overbar}[1]{\mkern 1.5mu\overline{\mkern-1.5mu#1\mkern-1.5mu}\mkern 1.5mu}

\newcommand{\be}{\begin{equation}}
\newcommand{\ee}{\end{equation}}
\newcommand{\ben}{\begin{equation*}}
\newcommand{\een}{\end{equation*}}
\newcommand{\ba}{\begin{eqnarray}}
\newcommand{\ea}{\end{eqnarray}}

\newcommand{\od}[1]{\begingroup\color[rgb]{0,0,0}#1\endgroup}

\newcommand{\ydpk}[1]{\begingroup\color[rgb]{0,0,0}#1\endgroup}

\graphicspath{{./figures/}}

\begin{document}

\title{Linear instability of turbulent channel flow}
\author{Pavan V. Kashyap}  %
\author{Yohann Duguet}
\affiliation{LIMSI-CNRS, UPR 3251, Universit\'e Paris-Saclay, 91405 Orsay, France}%
\author{Olivier Dauchot}
\affiliation{Gulliver Lab, UMR CNRS 7083, ESPCI Paris, PSL University, 75005 Paris, France}

\date{\today}

\begin{abstract}
Laminar-turbulent pattern formation is a distinctive feature of the intermittency regime in subcritical plane shear flows. \ydpk{By performing} extensive numerical simulations of the plane channel flow, we show that the pattern emerges from a spatial modulation of the turbulent flow, due to a linear instability. We sample over many realizations the linear response of the fluctuating turbulent field to a temporal impulse, in the regime where the turbulent flow is stable, just before the onset of the instability. The dispersion relation is constructed from the ensemble-averaged relaxation rates. As the instability threshold is approached, the relaxation rate of the least damped modes eventually reaches zero. The method allows, despite the presence of turbulent fluctuations and without any closure model, for an accurate estimation of the wavevector of the modulation at onset.
\end{abstract}

\maketitle
Turbulent channel flow is one of the most studied prototypes of inhomogeneous anisotropic turbulence.  It has been evidenced, both experimentally and numerically, that at moderate flow rates -- \ydpk{quantified} by the Reynolds number $Re$ -- it exhibits a spatio-temporally intermittent regime featuring robust large-scale turbulent structures amid a laminar background~\cite{tsukahara_dns_2005, hashimoto2009experimental, shimizu_bifurcations_2019}. The dynamical origin of such patterned turbulence in channel flow, as well as in other shear flows remains however actively debated~\cite{coles_transition_1965, atta_exploratory_1966, prigent_large-scale_2002, prigent_long-wavelength_2003, duguet_formation_2010, manneville2012turbulent,manneville2019subcritical,paranjape2020oblique,liu2021structured, klotz_phase_2022,kohyama2022sidewall}. 

On the lower end in $Re$ of the coexistence regime, turbulent patches grow and split, or decay, resulting in strongly fluctuating dynamics. It was suggested that the stochastic nature  of these processes, which decides whether turbulence will either spread or recede and eventually decay, could be described in the framework of non-equilibrium critical phenomena and specifically of directed percolation (DP), with the laminar state acting as the absorbing phase~\citep{pomeau_front_1986}. A major achievement of the past two decades has been to provide strong experimental and numerical evidences in favour of this scenario in a few shear flows~\cite{lemoult_directed_2016,chantry_universal_2017,klotz_phase_2022}. On the theoretical side, this regime has been described by an effective one--dimensional model of fronts in an excitable medium~\cite{barkley_simplifying_2011,barkley_modeling_2011}. Quite remarkably, and yet not theoretically understood, numerical simulations of the stochastic version of \ydpk{that} model reproduce the DP scenario.

Increasing $Re$, individual turbulent patches leave place to a coexistence organized at the flow scale~\citep{coles_transition_1965,atta_exploratory_1966}, in the form of a well organized periodic pattern of alternating laminar and turbulent bands, inclined at a well defined angle to the mean flow~\cite{prigent_large-scale_2002}. Considering the proliferation of turbulence as a problem of front propagation, it is tempting to view this pattern as packed arrays of individual localized structures. 
Yet, periodic pattern solutions have not been identified as solutions to the effective excitable dynamics~\cite{barkley_simplifying_2011,barkley_modeling_2011}. 
An alternative viewpoint is to consider the pattern as emerging from the featureless turbulence found at larger $Re$. Pioneering studies have demonstrated experimentally that the pattern developing in pCf and TCf, \ydpk{characterised} by two competing orientation of alternate sign, is fully captured by the dynamics of two coupled Ginburg-Landau equations with noise~\citep{prigent_large-scale_2002,prigent_long-wavelength_2003}. Recent visualizations, obtained in well-resolved numerical simulations of large domains channel flows, unveil small-amplitude harmonic modulations of the turbulent flow for values of $Re$ larger than those at which genuine laminar--turbulent coexistence is reported~\cite{shimizu_bifurcations_2019}. Statistical signatures of low-wall-shear-rate intermittency have been found, at $Re$-values usually associated with featureless turbulent flows~\citep{kashyap_flow_2020}. Altogether these results suggest the possibility of a large-wavelength instability of the \emph{turbulent flow} itself, as already proposed in~\citep{prigent_large-scale_2002}. As recently suggested on the basis of a spatiotemporal extension of a classical self-sustained turbulence model~\cite{waleffe1997self,dauchot2000phase}, the instability could be of Turing type~\citep{manneville2012turbulent}. However, there is no theoretical evidence for such a linear instability of the turbulent mean flow obtained from one-point closure models~\citep{schlatter_instability_2010}. These different viewpoints reflect global doubts regarding the origins of the patterned state and ambiguity as to whether the starting point for modeling the periodic pattern should be the spatial organization of the isolated turbulent patches when increasing $Re$, or the linear instability of the turbulent flow, including its fluctuations, when decreasing $Re$. 

Here we bring direct evidence in favour of the linear instability scenario in the case of the channel flow. To do so, we perform extensive numerical simulations and sample the linear response of the turbulent flow to a temporal impulse, in the regime where the large wavelength modulations are damped. The dispersion relation is then constructed from the ensemble-averaged relaxation rates, for decreasing values of $Re$. \ydpk{The smallest relaxation rate approaches zero for some critical value} $Re_c$, pointing at the spatial structure of the modes which grow at the instability onset. The method can be seen as the temporal counterpart of the spatial linear response considered in~\cite{russo2016linear}. It is intrinsically statistical in the sense that it establishes an {\it average dispersion relation} for the instability modes, from which the quantitative onset for the spatial modulation can be identified.
\begin{figure}
\begin{center}
\includegraphics[width=\columnwidth]{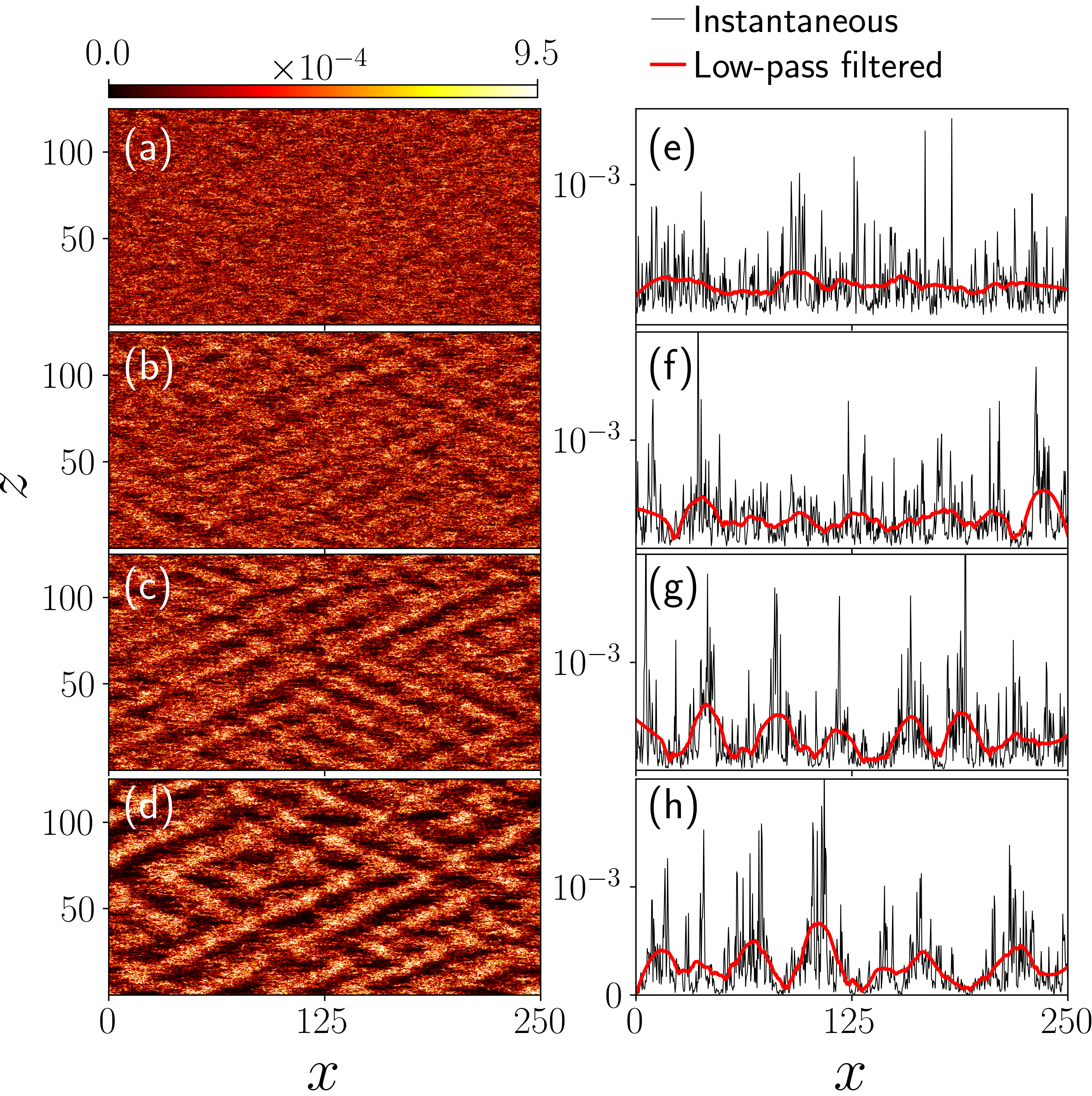}
\end{center}
\vspace{-5mm}
\caption{Onset of modulated turbulent channel flow. Left column: turbulent kinetic energy $E_v(x,z,t)$; right column: instantaneous streamwise profile of $E_v(x,z=cst,t)$, (black: raw instantaneous values, red: low-pass filtered values). From top to bottom : (a,e) $Re_\tau=100$, (b,f) $Re_\tau=90$, (c,g) $Re_\tau=85$, (d,h) $Re_\tau=80$.}
\label{bigpic}
\vspace{-5mm}
\end{figure}

\ydpk{The incompressible flow considered in this study is driven in the streamwise direction $x$ by a constant pressure gradient. The other Cartesian coordinates $y$ and $z$ are respectively wall-normal and spanwise. All length scales are nondimensionalised by the channel half-gap $h$, and velocities $\mathbf{u}=(u_x,u_y,u_z)$ by $U_{cl}$\ydpk{,} the centerline velocity of the classical laminar plane Poiseuille flow $U(y)=1-y^2$ driven by the same pressure gradient. Time is reported in units of $h/U_{cl}$. The velocity field is decomposed as $\mathbf{u}=U(y) \mathbf{e}_x + \mathbf{u}^\prime$ where $\mathbf{u}^\prime$ denotes the perturbation to the laminar base flow. Spatial averages are indicated with $\langle \sbullet \rangle_{x,y,z}$ where the subscript indicates the direction over which the average is computed. Time averages are indicated \ydpk{by} $\bar{\sbullet}$. Ensemble averages are indicated as $\langle \sbullet \rangle_e$. Fourier amplitudes are denoted with $\hat{\sbullet}$. The selected control parameter is the friction Reynolds number $Re_{\tau}=u_{\tau}h/\nu$, where $\nu$ is the kinematic viscosity of the fluid,  $u_{\tau}=\sqrt{\langle \overbar{\tau} \rangle _{xz}/\rho}$ is the friction velocity, with $ \langle \overbar{\tau} \rangle _{xz}$ the mean shear rate fixed by the pressure gradient, and $\rho$ the fluid density. Turbulent simulations were performed with the spectral solver Channelflow2.0 \citep{channelflow2} in a domain of $L_x=2L_z=$250 for times up to $t=4,000$.  These simulations are \ydpk{resolved} with a resolution of $N_x=N_z=$1024 (including dealiasing with the 2/3 rule) \od{and $N_y=65$} comparable to  \cite{shimizu_bifurcations_2019}.} The most recent investigations have reported laminar-turbulent patterns for $50\lesssim Re_{\tau}\lesssim 90$, and independent turbulent bands for lower values of $Re_{\tau}$ down to $\approx$ 36 \cite{shimizu_bifurcations_2019,kashyap_flow_2020,song_trigger_2020,mukund2021aging}. 

Large-scale modulations close to $Re_\tau \approx 90$, as well as genuine laminar-turbulent patterning for $Re_\tau \lesssim 90$, are unambiguous from Fig.~\ref{bigpic}, which displays the \od{instantaneous kinetic energy in the wall-normal direction} 
\begin{equation}
%E_v(x,z,t)= \langle {u^\prime}_y^2 \rangle_y ,\\
E_v(x,z,t)= \langle \frac{1}{2}{u_y^\prime}^2 \rangle_y,
\end{equation}
both at full spatial resolution and after application of a low-pass filter. It was checked that the modulations and the pattern are robust with respect to the doubling and quadrupling of the numerical domain in both $x$ and $z$ \od{(see Appendix-A of Supplementary material)}. The exact range of existence of the modulations, and notably their onset, are difficult to judge from visualizations alone because of the turbulent fluctuations, whose standard deviation can exceed the amplitude of the modulation. It is also sensitive to the choice of the visualized quantity.
\begin{figure}
\begin{center}
\includegraphics[width=\columnwidth]{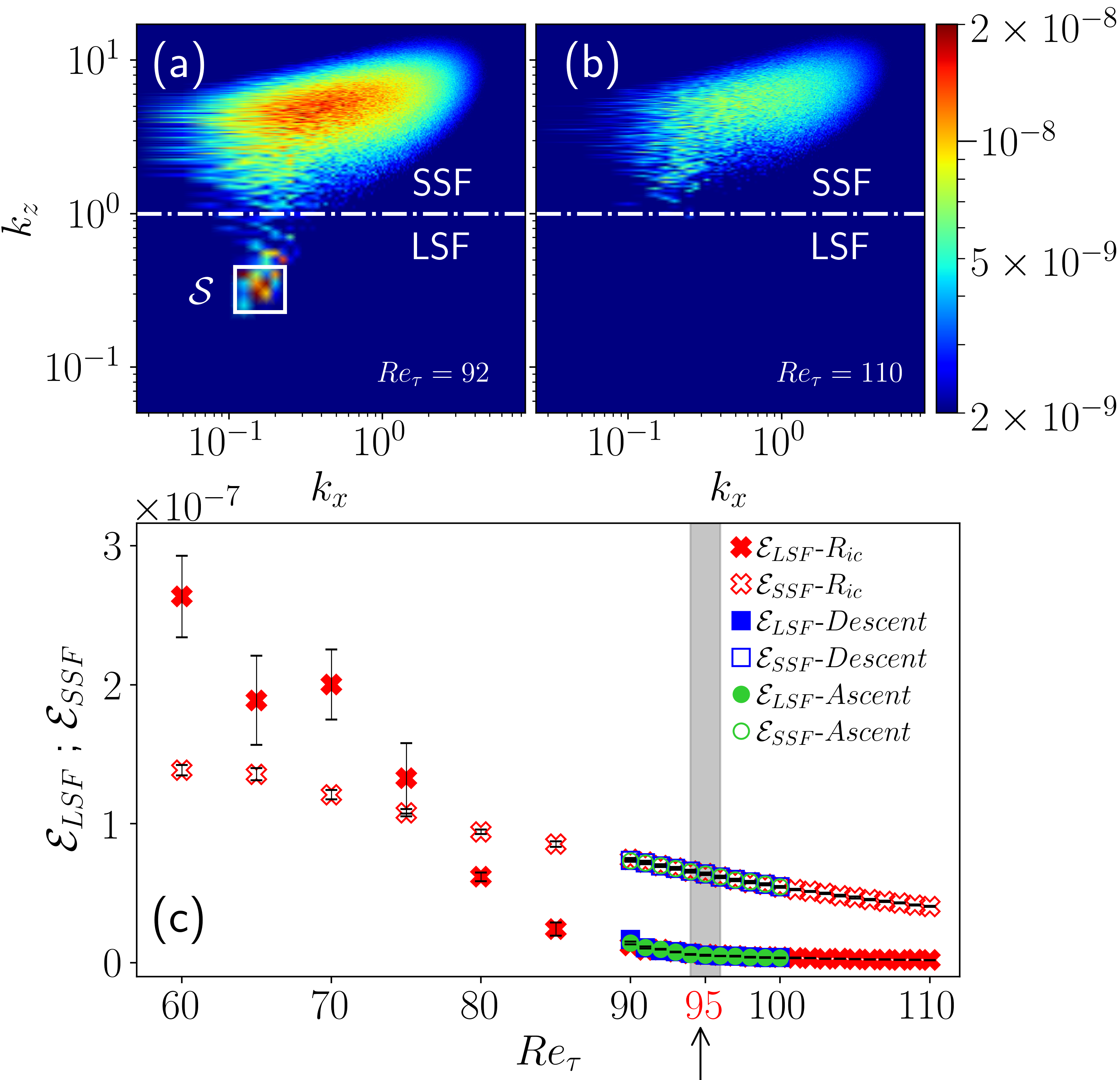}
\end{center}
\vspace{-5mm}
\caption{Premultiplied power spectrum for the $y$-averaged streamwise velocity fluctuation $\langle u_x^\prime \rangle_y$ at (a) $Re_\tau=92$ (b) $Re_\tau=110$ (right). (c) Energies $\mathcal{E}_{LSF}$ of the large-scale modes (defined according to Eq. \ref{amp_def}) and  $\mathcal{E}_{SSF}$ of the small scale modes vs. $Re_\tau$. The black arrow points at the linear instability threshold corresponding to the onset of the turbulent modulated flow determined here at $Re_{\tau}=95 \pm 1$. $\mathcal{E}_{LSF}$ \ydpk{(filled symbols) and $\mathcal{E}_{SSF}$ (empty symbols)} are computed for different simulation strategies - red cross : $R_{IC}$, divergence free random velocity field as initial condition, blue square : descending annealing, green circle : ascending annealing.}
\vspace{-5mm}
\label{fourier_spec}
\end{figure} 
\od{Conversely, the emergence of large-scale patterns, as $Re_{\tau}$ decreases, appears  clearly as a low-wavenumber signature in the time-averaged two-dimensional energy spectrum of the $y$-averaged fluctuating streamwise component $\langle u_x^\prime \rangle_y$ (Figure~\ref{fourier_spec}-a). Apart from the small-scale modes, corresponding to the turbulent fluctuations, one clearly observes a set of large-scale modes excited at $Re_\tau = 92$ (but absent at $Re_\tau = 110$). We also note an increase of the energy contained in the small--scale modes, and in the modes separating them from the large-scale ones, as $Re_{\tau}$ decreases. This highlights the persistent role of nonlinear triadic interactions between different scales. Still, two maxima are readily identified.} 

\ydpk{Exploiting this scale separation, we define $\mathcal{E}_{LSF}(Re_{\tau})$, and $\mathcal{E}_{SSF}(Re_{\tau})$, the dimensionless amplitude of the large-scale and small-scale flows, as the energy content of the spectral subdomains $\mathcal{S}_{LSF}=\{0 < k_z < 1 \}$ and $\mathcal{S}_{SSF}=\{1 \lesssim k_z \}$, respectively :
\begin{equation}
\mathcal{E}_{LSF/SSF}(Re_{\tau}) =\iint\limits_{\mathcal{S}_{LSF/SSF}} 
 \left| \langle \overbar{ \hat{u}_x^\prime } \rangle _y \right|^2 \, dk_x  \, d k_z. 
\label{amp_def}
\end{equation}
$\mathcal{E}_{LSF}$ is dominated by the low-$k$ modes inside the spectral subdomain $\mathcal{S}$ highlighted in Fig.~\ref{fourier_spec}-a(left). We checked that this scale separation, based solely on $k_z$,  appropriately delineates the two energy peaks observed in the spectra of all the turbulent fields we analysed.}
\od{$\mathcal{E}_{LSF}$ and $\mathcal{E}_{SSF}$ are shown as functions of $Re_{\tau}$ in Fig.~\ref{fourier_spec}-c obtained from random initial conditions ($R_{IC}$) or during slow ascent, respectively descent, annealing in $Re_{\tau}$. One observes a clear increase of $\mathcal{E}_{LSF}$ in contrast with the marginal increase of $\mathcal{E}_{SSF}$ as $Re_{\tau}$ decreases from the featureless turbulent regime ($Re_{\tau}\gtrsim 110$) to the well-defined pattern one ($50\lesssim Re_{\tau}\lesssim 90$), with no sign of hysteresis. We note that $\mathcal{E}_{LSF}$ is never strictly zero even at high $Re_{\tau}$. Whether the above observations result from a true bifurcation or are simply a mere crossover cannot be decided by simply looking at Figure 2. This is what motivates the following analysis where we show that the rise of $\mathcal{E}_{LSF}$ is due to a linear instability of the turbulent flow. }

Establishing the linear instability of a flow with arbitrary time-dependence can be addressed in different ways. One possibility is to study the linear stability of the {\it mean} flow using the Orr-Sommerfeld formalism. This strategy, whether conducted at high~\cite{reynolds1967stability} or transitional~\cite{tuckerman_private} $Re$, predicts linear stability. At the opposite end, taking into account all temporal fluctuations is in principle possible using Lyapunov analysis. However for turbulent flows the number of positive Lyapunov exponents is prohibitively huge~\cite{keefe1992dimension} because of the chaoticity at small scales down to the Kolmogorov scale. The turbulent scales where these instabilities dominate are however {\it not} the emerging large-scales visible in Fig.~\ref{fourier_spec}, \od{which suggests the computation of  alternative quantities.}

The general idea is to study the linear response of the flow to a temporal impulse. If the flow is linearly stable, the disturbance should relax, otherwise it should grow and lead to a bifurcated flow. However, the reference flow being turbulent, the analysis must be conducted at a statistical level. \ydpk{Besides} the spatial structure of the temporal impulse should be agnostic to the turbulent spectrum. We therefore proceed as follows.
A representative turbulent state in the statistically steady regime at the required value of $Re_{\tau}$, simulated for $t<0$, is perturbed at $t=0$ using a divergence-free \emph{noise field}, before the simulations runs further without noise, for $t>0$, and we monitor the temporal evolution of the modulus of the Fourier amplitudes of {\it large-scale modes}, \ydpk{$\left| \hat{ \tau } \right| (Re_{\tau}, k_x, k_z, t)$}, with $(k_x, k_z) \in \mathcal{S}=\{0.075 \lesssim k_x \lesssim 0.22, 0.2 \lesssim k_z \lesssim 0.5 \}$ (the highlighted square area in Fig.~\ref{fourier_spec}-a, part of $\mathcal{S}_{LSF}$).  
\begin{figure}[t!]
\begin{center}
\includegraphics[width=\columnwidth]{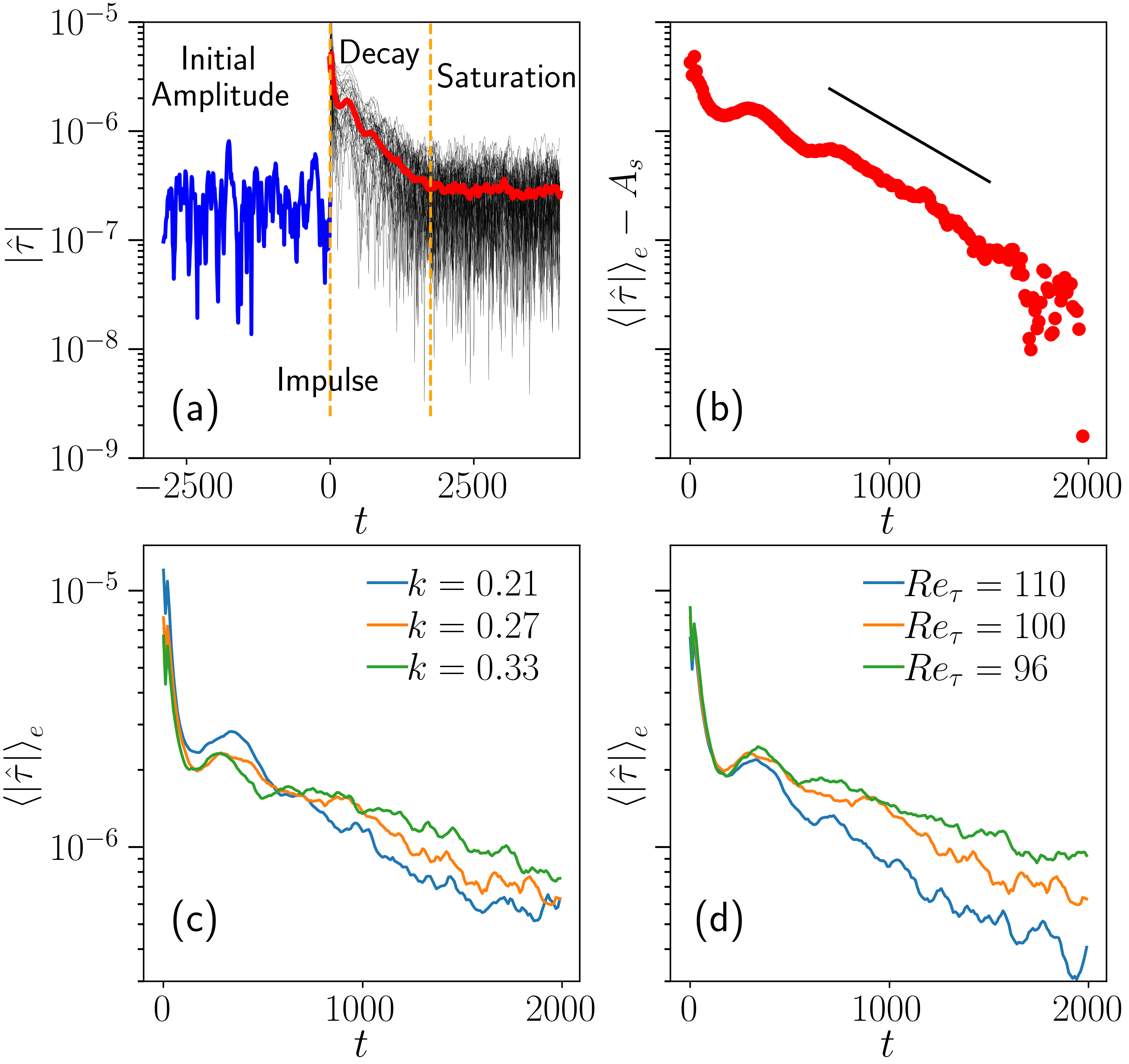}
\end{center}
\caption{(a) Temporal evolution of the amplitude of the Fourier mode $ \left| \hat{ \tau } \right|  (Re_\tau, k_x, k_z, t)$ with $(k_x, k_z) = (0.12, 0.3)$, ($k=0.32$) and $Re_\tau=120$; blue $(t<0)$, black: individual realizations for $t>0$ and red: ensemble average $\langle \left| \hat{ \tau } \right|  \rangle _e (t)$. (b) $ \langle \left| \hat{ \tau } \right|  \rangle _e (t) - A_s $, same $(k_x, k_z)$, same $Re_{\tau}$, black: exponential fit. (c) $\langle \left| \hat{ \tau } \right|  \rangle _e (t)$ for three different values of $k$, $Re_{\tau}=100$. (d) $\langle \left| \hat{ \tau } \right|  \rangle _e (t)$ for three different values of $Re_{\tau}$, $k=0.27$.}
\label{impulse}
\end{figure}
For large enough $Re_\tau$, the disturbed flow relaxes back towards the steady turbulent state. The individual time series $\left| \hat{ \tau } \right| $ however showcase a strongly fluctuating decay. This computational decay experiment is therefore repeated over $40$ different realizations of the noise field and the modulus of the spectral amplitude of each large-scale Fourier mode, \ydpk{$|\hat{\tau}|$,} is ensemble-averaged over all realizations to yield $\langle \left| \hat{ \tau } \right|  \rangle _e$, as illustrated in figure~\ref{impulse}-a for $Re_\tau=120$ and $(k_x,k_z)=(0.12,0.3)$. Ensemble-averaging brings clarity into the system's response: past an initially nonlinear decrease of $\langle \left| \hat{ \tau } \right|  \rangle _e$, a clear exponential decay towards a finite value $A_s$ is observed. \od{Given that this exponential decay is relevant only at the late stage of the relaxation, changing the amplitude of the initial noise field is not relevant}. This exponential decay captures the \textit{averaged linear response} of the turbulent state with respect to a temporal impulse.  The corresponding growth rate $\sigma$ is evaluated by estimating first the saturation level $A_s$ and then fitting an exponential decay to $(\langle \left| \hat{ \tau } \right| \rangle _e - A_s)$, using a straight line fit in logarithmic scale, as portrayed in figure~\ref{impulse}-b (See Appendix B of Supplementary material for a detailed step by step description of the procedure). As a first step, the analysis is carried out along the diagonal of the spectral window $\mathcal{S}$. Fig.~\ref{impulse}-c,d show the strong dependence of the growth rate on both  $k=\sqrt{k_x^2 + k_z^2}$ and $Re_{\tau}$.
More specifically, one observes that, for $Re_{\tau} = 96$, the growth rate of the mode corresponding to $k=0.38$ is  close to vanishing, suggesting the proximity of a linear instability. In principle one could expect monitoring the average exponential growth of such a large-scale mode beyond the instability threshold. \ydpk{However, not only would the growth rate be hard to measure accurately near onset, one would also need to isolate the featureless turbulent state in a regime where it is unstable.}

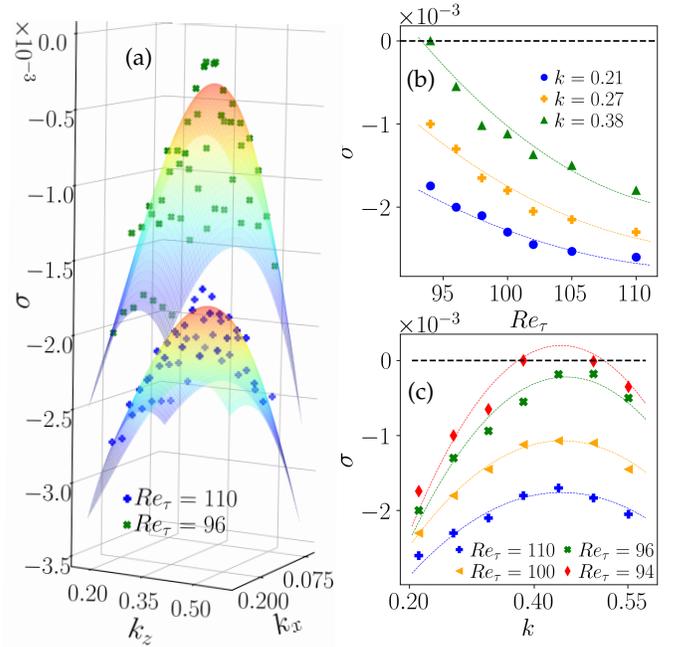
\begin{figure}[t!]
\begin{center}
\begin{tikzpicture}
%Position the 2D dispersion plot
\node[inner sep=0pt] (f1) at (3.4,0){\includegraphics[scale=0.42]{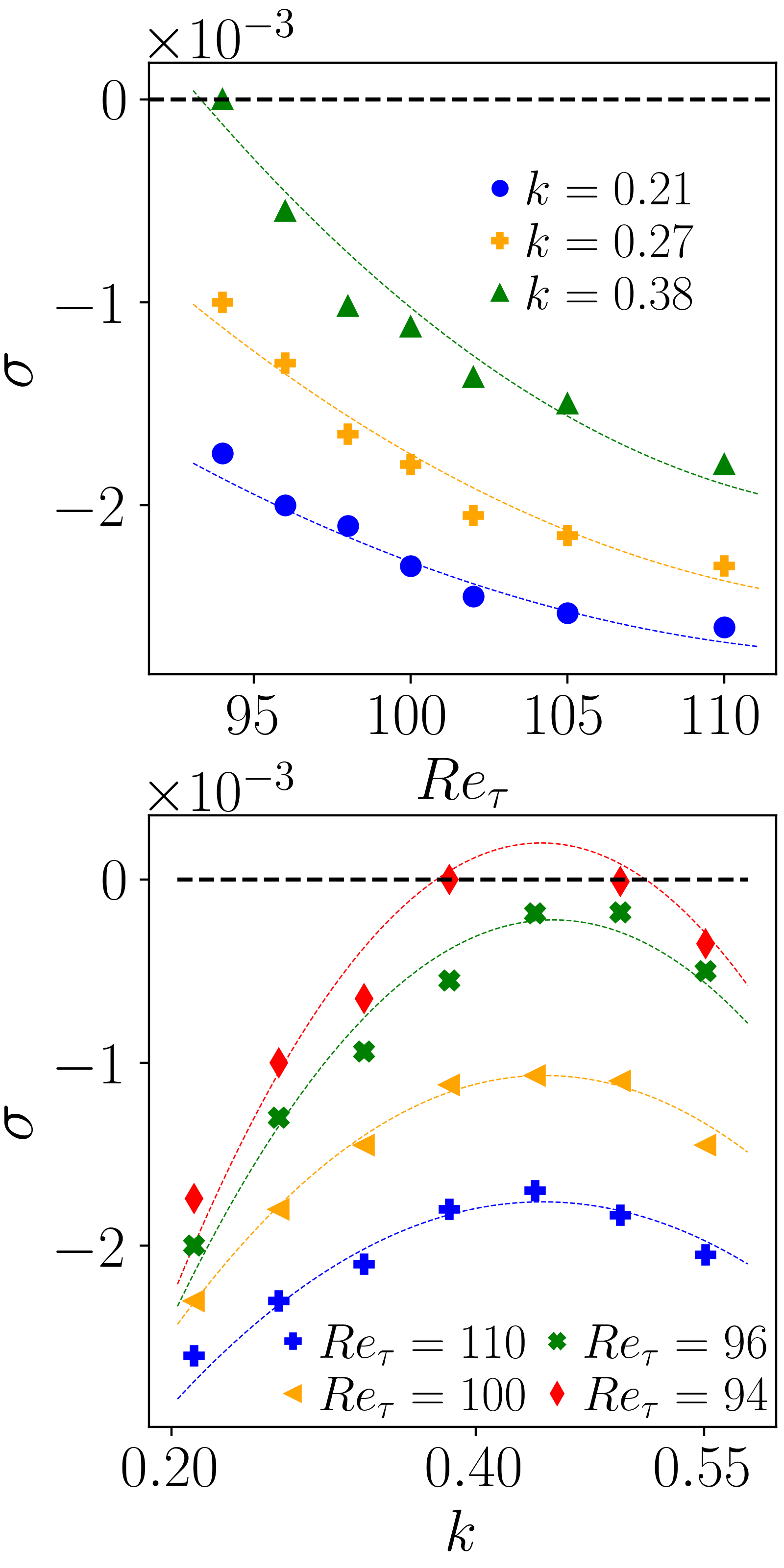}};
\node at (2.35,3.2) {(b)};
\node at (2.35,-0.95) {(c)};
%%
%Position the 3D plot
\node[inner sep=0pt] (f2) at (-0.95,-0.2){\includegraphics[scale=0.51,angle=-1]{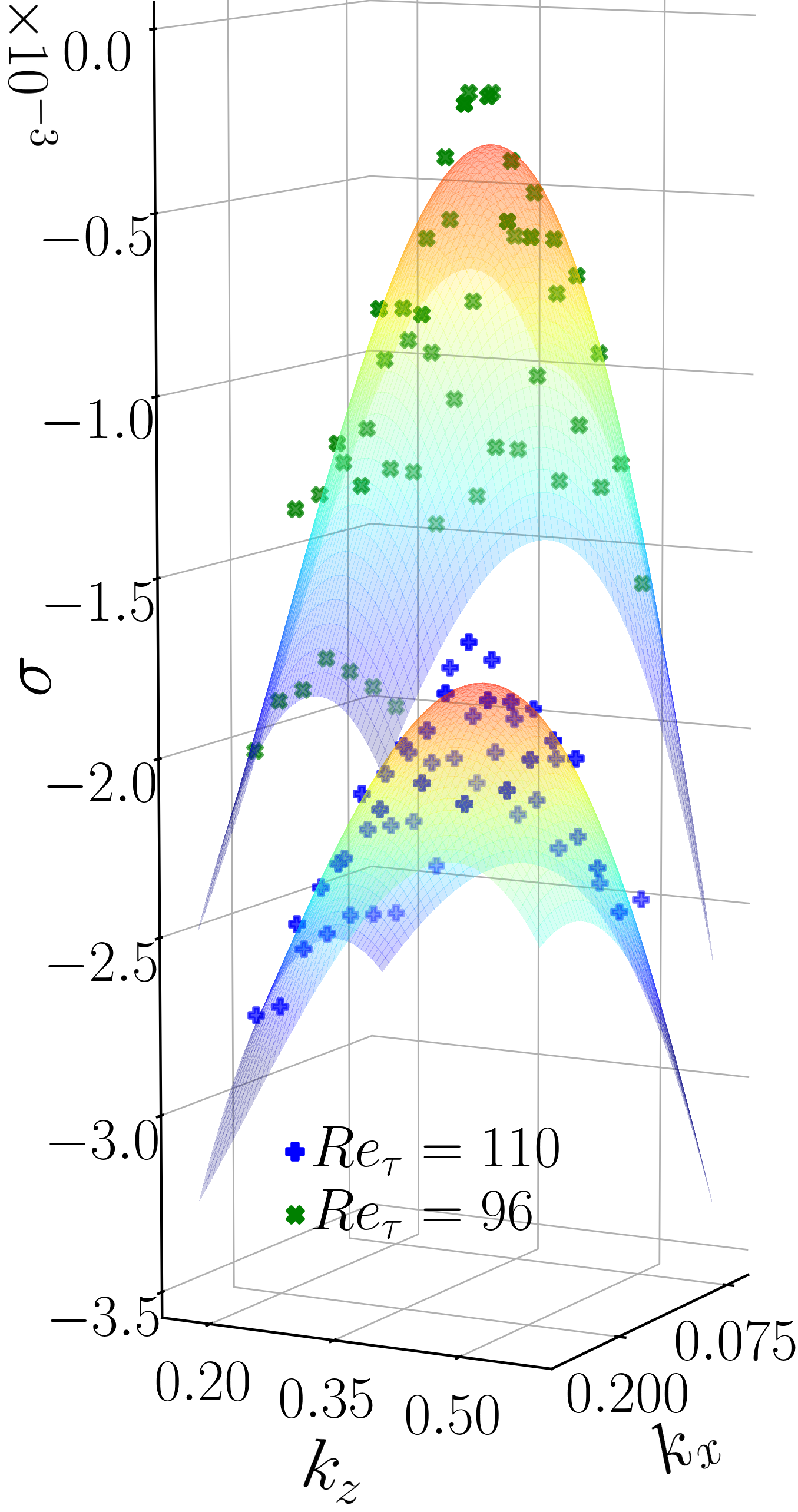}};
\node at (-1.4,3.5) {(a)};
\end{tikzpicture}
\end{center}
\caption{Dispersion relation: (a) growth rate $\sigma$ versus both $k_x$ and $k_z$ for two values of $Re_{\tau}$=96 and 110. (b) growth rate $\sigma$ versus $Re_\tau$ parametrized by the wavenumber modulus $k=\sqrt{k_x^2+k_x^2}$ (c) growth rate $\sigma$ versus the wavenumber modulus $k$ parametrized by $Re_\tau$ \od{(The data in (a) is fitted with a paraboloid surface while in (b) and (c) the continuous curves are guides to the eye).} 
\vspace{-5mm}
}
\label{dispersion}
\end{figure}

We therefore concentrate on the decay rates and extract the mean dispersion relation for the linear response of the turbulent flow (Fig.~\ref{dispersion}). \ydpk{The data is fit with a paraboloid surface (Fig.~\ref{dispersion}a) of the form :
\begin{equation}
\sigma = \alpha (k_x -k_{xc})^2 + \beta (k_z - k_{zc})^2 +\gamma (k_x - k_{xc})(k_z - k_{zc}) + \delta ,   \label{pbfit} 
\end{equation}
\od{where $(k_{xc}, k_{zc})$ is the critical wavevector.} The coefficients of Eq~\ref{pbfit} obtained for different values of $Re_\tau$ are \ydpk{reported} in the supplementary material Appendix B.}
The dispersion curves approach the neutral axis as the value of $Re_{\tau}$ is decreased and eventually cross it for $Re_{\tau}=94$. The estimated critical value for the instability is $Re_{\tau} = 95 \pm 1$. The critical wavevector $(k_{xc}=0.18 \pm 0.025,k_{zc}=0.42 \pm 0.05)$ is obtained from the above parabolic fit of the decay rates, estimated for all $(k_x,k_z)\in\mathcal{S}$, as illustrated in Fig.~\ref{dispersion}-a for $Re_\tau=110$ and $Re_\tau=96$. It perfectly matches the one measured directly at onset and lead to an inclination of the pattern with the streamwise direction of $23\pm 0.5^{\circ}$, consistently with the measurements reported in~\cite{kashyap_flow_2020}.
This quantitative agreement validates the proposed methodology, i.e. the statistical analysis of the temporal impulse response can be considered as a new experimental/numerical method to address the linear stability analysis of a steady, but fluctuating dynamics. \ydpk{We emphasize again} that the base flow for the analysis is the turbulent flow itself, including all fluctuations~\cite{iyer_identifying_2019}, not the mean flow.

Altogether our results provide direct evidence for a linear instability of the turbulent state itself, as first conjectured in Ref.~\cite{prigent_large-scale_2002}. This linear instability leads to a spatial modulation of the turbulent flow, the amplitude of which grows and saturates according to weakly non-linear contributions~\cite{prigent_long-wavelength_2003}. 
For low enough $Re_\tau$, the modulation breaks into a pattern of alternated turbulent and laminar bands. Further decreasing $Re_\tau$ these bands gain in independence and a proper stochastic front dynamics sets in. 

Our work paves the way for future works in two main directions. First, one would like to identify the instability mechanism. A possible candidate, commonly encountered across diverse noisy chemical and biological systems \cite{cross_pattern_2009} relies on the Turing instability \cite{manneville2012turbulent,kashyap2021subcritical}.  
It is based on the competition between an inhibitor and an activator field with different diffusivities~\cite{turing_chemical_1952}. However this approach requires modelling of the turbulent diffusivity using e.g. simple first-moment closures \citep{reynolds1972mechanics,del2006linear}.
Another possible approach is to consider a generalised stability analysis taking into account higher-order moments of the fluctuations \citep{markeviciute2022improved}. Both approaches are based on closure assumptions. The instability unveiled in the present work represents an ideal and simple case to test these assumptions. 
The second future direction of research consists in identifying the strongly nonlinear scenario along which the pattern looses its spatial coherence. It remains a formidable challenge.

\acknowledgements{{\it Acknowledgements:} This study was made possible using computational resources from IDRIS (Institut du D\'eveloppement et des Ressources en Informatique Scientifique) and the support of its staff. The developing team of channelflow.ch is also gratefully thanked. The authors also acknowledge constructive discussions with D. Barkley, L.S. Tuckerman, S. Gom\'e and P. Manneville.}

\bibliographystyle{apsrev_nurl}
\bibliography{References}

\section{Supplementary Informations}
\subsection{Appendix A : Illustration of pattern formation for different domain sizes}
\begin{figure}[h!]
    \centering
    \includegraphics[width=0.95\columnwidth]{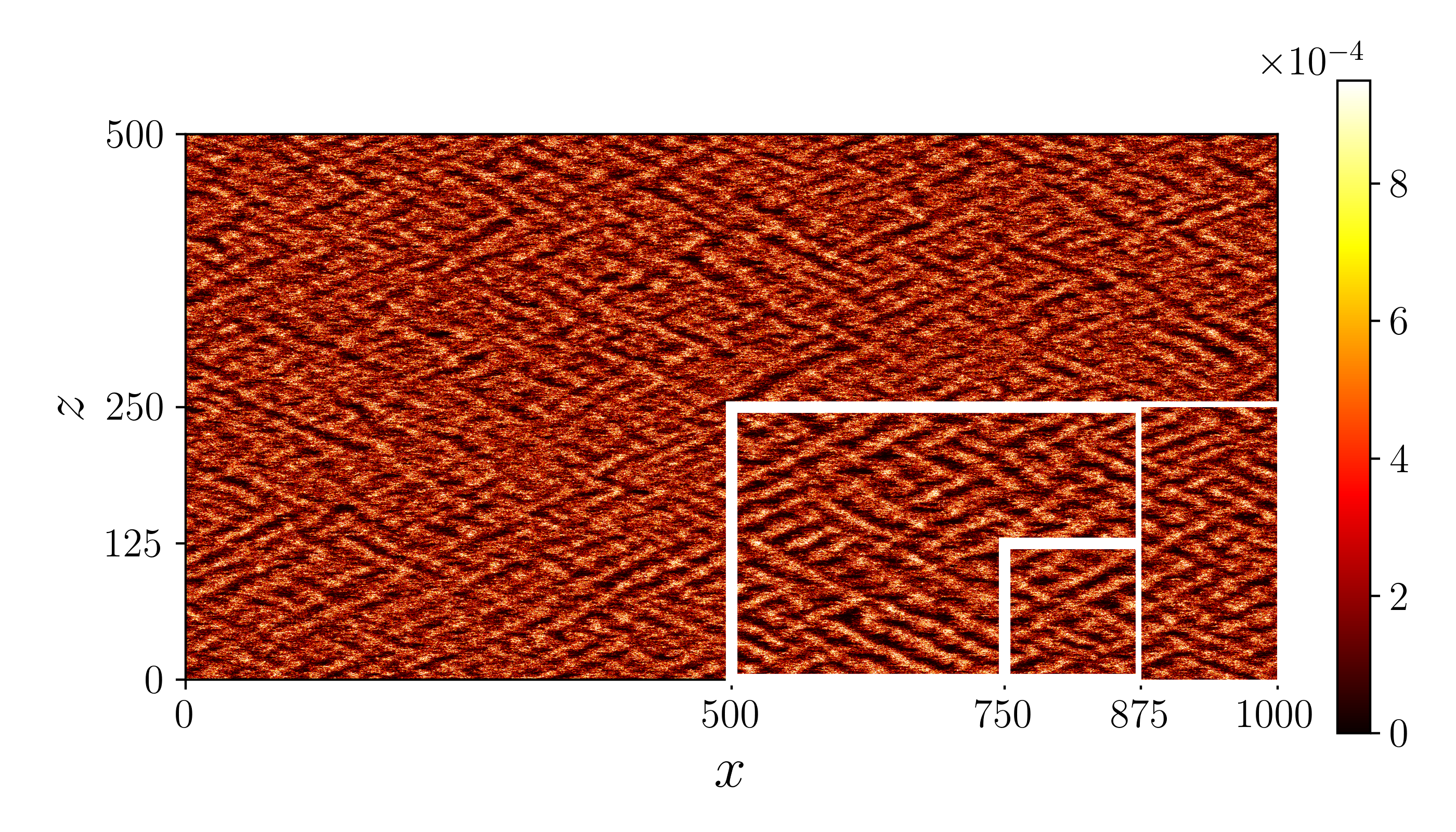}
    \caption{Illustration of pattern formation in different spatial domains. $E_v(x, z)$ for $Re_\tau = 80$, same colorbar as fig. 1 of the main article. Each domain demarcated by white lines corresponds to a simulation in a different computational domain. }
    \label{ptil}
\end{figure}

The robustness of pattern formation has been verified with simulations conducted for different domain sizes with equivalent numerical resolutions. As an example, fig~\ref{ptil} illustrates pattern formation occurring at $Re_\tau=80$, visualized with the help of $E_v(x,z)$ for different domain sizes i.e $L_x=2L_z=250$, $L_x=2L_z=500$, $L_x=2L_z=1000$. It demonstrates the robustness of the wavelength with respect to changes of computational domain. Simulations with the inverse aspect ratio, i.e $2L_x=L_z=250$, confirm this observation. 

\subsection{Appendix B : Evaluation of the decay rate $\sigma$}
The procedure followed for evaluating the decay rate is as follows:
\begin{itemize}
    \item Plot the ensemble-averaged amplitude $\langle \left| \hat{\tau} \right| \rangle _e (Re_\tau,k_x,k_z,t)$ against time.
    \item Extract the saturation amplitude : The largest time interval of least change in the amplitude indicates the statistically steady state. The average computed over this time interval is denoted as the saturation amplitude $A_s$.
    \item The saturation amplitude $A_s$ is subtracted from $\langle \left| \hat{\tau} \right| \rangle _e (Re_\tau,k_x,k_z,t)$
    \item The logarithm of the amplitude $\langle \left| \hat{\tau} \right| \rangle _e - A_s$ is fit with a straight line by considering $\log(\langle \left| \hat{\tau} \right| \rangle _e - A_s)$ for the time interval $(t,t_s)$ where $t_s$ marks the beginning of saturation and $t$ is gradually increased from zero. Since the decay / approach to the steady state is exponential as observed from fig~3b, the logarithmic scale makes it linear in nature. Performing the linear curve fit for different intervals of $(t,t_s)$, the largest interval of marginal change in the slope of the fitted line is taken to be representative of the exponential decay. The slope thus evaluated is considered as the decay rate $\sigma(Re_\tau,k_x,k_z)$
    \item This procedure is repeated for all values of $Re_\tau$ simulated and $(k_x,k_z)$ pairs in the interval $\mathcal{S}=\{0.075 \lesssim k_x \lesssim 0.22, 0.2 \lesssim k_z \lesssim 0.5 \}$ to construct the dispersion surfaces shown in fig~4a.
\end{itemize}

Following the evaluation of the decay rates $\sigma(Re_\tau,k_x,k_z)$, the data is fit with a paraboloid for the dispersion surface (see fig~4a). The functional form of these fits and the obtained coefficients are listed below:

\begin{itemize}
    \item The dispersion surface is fit with the function 
    
    \begin{equation}
    \sigma= \alpha(k_x-k_{xc})^2 + \beta(k_z - k_{zc})^2 + \gamma (k_x - k_{xc})(k_z - k_{zc}) + \delta     \label{paraboloid}
    \end{equation}
    
    The coefficients of the fits are :
    
    \begin{table}[h!]
        \centering
        \begin{tabular}{|c|c|c|c|c|c|c|}
            \hline
            \rule{0pt}{3.2ex} $Re_\tau$ & $\alpha$ & $\beta$ & $\gamma$ & $\delta$ & $k_{xc}$ & $k_{zc}$ \\[2ex]
            \hline
            \rule{0pt}{3.2ex} 110 & -0.069 & -0.01 & 0.018 & -0.0018 & 0.18 & 0.4 \\[2ex]
            \hline
            \rule{0pt}{3.2ex} 96 & -0.1 & -0.016 & 0.028 & -0.0003 & 0.19 & 0.45 \\[2ex]
            \hline
        \end{tabular}
        \caption{Coefficients of the paraboloid surface fit of Eq~\ref{paraboloid}}
        \label{csurf}
    \end{table}

\end{itemize}

\end{document}